\PassOptionsToPackage{dvipsnames}{xcolor}
\documentclass[%
  aip,%
 amsmath,amssymb,floatfix,
reprint,%
twocolumn,%
jcp,%
longbibliography,
]{revtex4-1} 

\usepackage[normalem]{ulem}
\usepackage[all]{xy}
\usepackage{xcolor}

\usepackage{amsmath,amsthm,latexsym,amssymb,amsfonts,epsfig}

\usepackage[english]{babel}
\usepackage[utf8]{inputenc}			
\usepackage{graphicx, texdraw}  
\usepackage{color}
\usepackage{bm}

\usepackage{type1ec}

\usepackage{mathtools}
\usepackage{mathrsfs} 

\usepackage[left=1.3cm, right=1.3cm, top=2.cm, bottom=2.cm]{geometry}  

\usepackage{enumerate}

\usepackage[colorlinks=true,citecolor=OliveGreen, linkcolor=blue]{hyperref}

\usepackage{csquotes} 

\usepackage{color}

\usepackage{amsmath}
\usepackage{amsfonts}
\usepackage{amssymb}

\usepackage{epstopdf}

\newcommand{\vect}[1]{\bm{#1}}

\newcommand{\bv}{\vect{v}}

\newcommand{\G}{\mathcal{G}}
\newcommand{\R}{\vect{r}}
\newcommand{\Intd}{\mathrm{d }}

\newcommand{\F}{\vect{F}}

\newcommand{\bigO}{\mathcal{O}}

\newcommand{\Faxen}{Fax\'{e}n}

\newcommand{\bNabla}{\boldsymbol{\nabla}}

\newcommand{\BF}{\boldsymbol{F}}

\newcommand{\BL}{\boldsymbol{L}}

\newcommand{\BV}{\boldsymbol{V}}

\newcommand{\mi}{\bm{\mu}}

\newcommand{\zP}{z_\mathrm{P}}

\newcommand{\zT}{z_\mathrm{T}}

\newcommand{\psiT}{\psi_\mathrm{T}}

\usepackage{setspace}

\begin{document}

\title{Swimming trajectories of a three-sphere microswimmer near a wall}

\author{Abdallah Daddi-Moussa-Ider}
\email{ider@thphy.uni-duesseldorf.de}

\affiliation
{Institut f\"{u}r Theoretische Physik II: Weiche Materie, Heinrich-Heine-Universit\"{a}t D\"{u}sseldorf, Universit\"{a}tsstra\ss e 1, D\"{u}sseldorf 40225, Germany}

\author{Maciej Lisicki}

\affiliation
{Department of Applied Mathematics and Theoretical Physics, University of  Cambridge, Wilberforce Rd, Cambridge CB3 0WA, United Kingdom}

\affiliation
{Institute of Theoretical Physics, Faculty of Physics, University of Warsaw, Pasteura 5, 02-093 Warsaw, Poland }

\author{Christian Hoell}

\affiliation
{Institut f\"{u}r Theoretische Physik II: Weiche Materie, Heinrich-Heine-Universit\"{a}t D\"{u}sseldorf, Universit\"{a}tsstra\ss e 1, D\"{u}sseldorf 40225, Germany}

\author{Hartmut Löwen}
\email{hlowen@thphy.uni-duesseldorf.de}

\affiliation
{Institut f\"{u}r Theoretische Physik II: Weiche Materie, Heinrich-Heine-Universit\"{a}t D\"{u}sseldorf, Universit\"{a}tsstra\ss e 1, D\"{u}sseldorf 40225, Germany}

\begin{abstract}

The hydrodynamic flow field generated by self-propelled active particles and swimming microorganisms is strongly altered by the presence of nearby boundaries in a viscous flow. Using a simple model three-linked sphere swimmer, we show that the swimming trajectories near a no-slip wall reveal various scenarios of motion depending on the initial orientation and the distance separating the swimmer from the wall. 
We find that the swimmer can either be trapped by the wall, completely escape, or perform an oscillatory gliding motion at a constant mean height above the wall. Using a far-field approximation, we find that, at leading order, the wall-induced correction has a source-dipolar or quadrupolar flow structure where the translational and angular velocities of the swimmer decay as inverse third and fourth powers with distance from the wall, respectively. The resulting equations of motion for the trajectories and the relevant order parameters fully characterize the transition between the states and allow for an accurate description of the swimming behavior near a wall. 
We demonstrate that the transition between the trapping and oscillatory gliding states is first order discontinuous, whereas the transition between the trapping and escaping states is continuous, characterized by non-trivial scaling exponents of the order parameters.
In order to model the circular motion of flagellated bacteria near solid interfaces, we further assume that the spheres can undergo rotational motion around the swimming axis. 
We show that the general three-dimensional motion can be mapped onto a quasi-two-dimensional representational model by an appropriate redefinition of the order parameters governing the transition between the swimming states.

\end{abstract}
\date{\today}

\maketitle

\setstretch{1.}

\section{Introduction}

Swimming microorganisms use a variety of strategies to achieve propulsion or stir the suspending fluid \cite{brennen77}. To circumvent the constraint of time reversibility of the Stokes equation governing the small-scale motion of a viscous fluid, known as Purcell's scallop theorem \cite{purcell77}, many of them rely on the non-reciprocal motion of their bodies. To understand the nature of this process, a number of artificial designs have been proposed to construct and fabricate model swimmers capable of propelling themselves in a viscous fluid by internal actuation. Among these, a particular class are simplistic systems with only few degrees of freedom necessary to break kinematic reversibility, as opposed to continuous irreversible deformations or chemically-powered locomotion \cite{lauga09, marchetti13, elgeti15, bechinger16, zottl16, lauga2016ARFM}. A famous example of such a design is the swimmer of Najafi and Golestanian \cite{najafi04} encompassing three aligned spheres; their distances vary in time periodically with phase differences, thus leading to locomotion along straight trajectories \cite{Najafi2005,Golestanian2008a,Golestanian2008b, alexander09}. This system has been also realized experimentally using optical tweezers \cite{Leoni2009, grosjean16}. 
Notably, a number of similar designs have been proposed: with one of the arms being passive and elastic  \cite{Montino2015}, both arms being muscle-like \cite{Montino2017} or using a bead-spring swimmer model \cite{pande15, pande17, babel16}. Variations of this idea leading to rotational motion have been proposed: a circle swimmer in the form of three spheres joined by two links crossing at an angle \cite{Ledesma2012}, linked like spokes on a wheel \cite{Dreyfus2005} or connected in an equilateral triangular fashion ~\cite{rizvi18}. 
An extension to a collection of $N>3$ spheres has also been considered \cite{Felderhof2006}.
Further investigations include the effect of fluid viscoelasticity~\cite{pak12, zhu12, curtis13, yazdi14, yazdi15, datt15, yazdi17}, swimming near a fluid interface~\cite{pimponi16, dominguez16, dietrich17} or inside a channel~\cite{elgeti09, zhu13,yang16, elgeti16, liu16prl}, and the hydrodynamic interactions between two neighboring microswimmers near a wall\cite{li14}.
Intriguing collective behavior and spatiotemporal patterns may arise from the interaction of many swimmers, including the onset of propagating density waves~\cite{gregoire04, mishra10, heidenreich11, menzel12, liebchen16, goff16, liebchen17, liebchen17b} and laning~\cite{menzel13, kogler15, romanczuk16, menzel16}, the motility-induced phase separation~\cite{tailleur08, palacci13, buttinoni13, speck14, speck15} and the emergence of active turbulence~\cite{wensink12pnas, wensink12, dunkel13, heidenreich14, kaiser14, heidenreich16, lowen16}. Boundaries have also been shown to induce order in collective flows of bacterial suspensions \cite{woodhouse2012,wioland2013,wioland2016}, leading to potential applications in autonomous microfluidic systems \cite{woodhouse2017}. A step towards understanding these collective phenomena is to explore the dynamics of a single model swimmer interacting with a boundary.

The long-range nature of hydrodynamic interactions in low Reynolds number flows results in geometrical confinement significantly influencing the dynamics of suspended particles or organisms \cite{happel54}. Interfacial effects govern the design of microfluidic systems \cite{stone04, squires2005, lauga07}, they hinder translational and rotational diffusion of colloidal particles \cite{lisicki14, lisicki16, daddi16c, rallabandi17, rallabandi17b, daddi17, daddi17b, daddi17c, daddi17d}, and play an important role in living systems, where walls have been shown to qualitatively modify the trajectories of swimming {\it E. coli} bacteria~\cite{frymier95, diluzio05, berke08, drescher11, mino11, lushi17}  or microalgae~\cite{ishimoto13, contino15}. Simplistic two-sphere near-wall models of bacterial motion have revealed that the dynamics of a bead swimmer can be surprisingly rich, including circular motion in contact with the wall, swimming away from the wall, and a non-trivial steady circulation at a finite distance from the interface \cite{dunstan2012}. This diverse phase behavior has also been corroborated in systems of chemically powered autophoretic particles \cite{Uspal2015a,Uspal2015b, simmchen16, ibrahim15, mozaffari16, popescu16, ruhle17, mozaffari18}, leading to a phase diagram also includes trapping, escape, and a steady hovering state.
Swimming near a boundary has been addressed using a two-dimensional singularity model combined with a complex variable approach~\cite{crowdy11}, a resistive force theory~\cite{di-leonardo11}, and a multipole expansion technique~\cite{lopez14}.
It has further been demonstrated that geometric confinement can conveniently be utilized to steer active colloids along arbitrary trajectories~\cite{das15}.
The detention times of microswimmers trapped at solid surfaces has been studied theoretically, elucidating the interplay between hydrodynamic interactions and rotational noise~\cite{schaar15}. Trapping in more complex geometries has particularly been analyzed in the context of collisions of swimming microorganisms with large spherical obstacles~\cite{spagnolie2015, desai18} and scattering on colloidal particles \cite{shum2017}. The generic underlying mechanism is thought to play a role in a number of biological processes, such as the formation of biofilms~\cite{loosdrecht1990, drescher13}.

In order to analyze the dynamics of a neutral three-sphere model swimmer near a no-slip wall, Zargar {\it et al.} \cite{Zargar2009} calculated the phase diagram, finding that the swimmer always orients itself parallel to the wall. In their calculation, they expand the hydrodynamic forces in the small parameter $\epsilon=L/z$, where $L$ is the length of the swimmer and~$z$ is the wall-swimmer distance, arriving at the conclusion that the dominant term is proportional to $z^{-2}$. In this contribution, we revisit this problem and demonstrate that the dominant term in the swimming velocities scales rather as~$z^{-3}$. This allows us to calculate the full phase diagram which shares qualitative features seen in the aforementioned artificial microswimmers, that is steady gliding, trapping, and escaping trajectories, basing on the initial conditions of the swimmer. 
\bigskip

The paper is organized as follows.
In Sec.~\ref{section-theoretical-model}, we introduce a theoretical model for the swimmer and derive the governing equations of motion in the low-Reynolds-number regime.
We then present in Sec.~\ref{section-state-diagram} a state diagram of swimming near a hard wall and introduce suitable order parameters governing the transitions between the states.
In Sec.~\ref{section-far-field-model}, we present a far-field theory that describes the swimming dynamics in the limit far away from the wall.
We then discuss in Sec.~\ref{section-effect-rotation} the effect of the rotation of the spheres on the swimming trajectories and show that the general 3D motion can be mapped onto a 2D generic model by properly redefining the order parameters.
Finally, concluding remarks are contained in Sec.~\ref{section-conclusions}.

\allowdisplaybreaks

\section{Theoretical model}\label{section-theoretical-model}

\subsection{Stokes hydrodynamics}

We consider the (sufficiently slow) motion of a swimmer moving in the vicinity of an infinitely extended planar hard wall. Since systems of biological or microfluidic relevance are typically micrometer-sized, the Reynolds number is low, and the dynamics are dominated by viscosity. For small amplitude and frequency of motion, the fluid flow surrounding the swimmer is governed by the steady incompressible Stokes equations~\cite{kim13}, which for a point force acting on the fluid at position~$\R_0$ relate the velocity $\vect{v}$ and pressure field, $p$, by
\begin{align}
\eta \bNabla^2 \bv(\R) - \bNabla p(\R) + \vect{F} \delta(\R-\R_0) &= 0 \, , \label{stokesGleischung_1} \\
\bNabla \cdot \bv (\R) &= 0 \, ,  \label{stokesEq_2}
\end{align}
where $\eta$ denotes the dynamic viscosity of the fluid.

In an unbounded fluid, the solution of this set of equations for the velocity field is expressed in terms of the Green's function
\begin{equation}
	v_\alpha (\R) = \G_{\alpha\beta} (\R,\R_0) F_\beta \, ,  
\end{equation}
for $\alpha,\beta \in \{x,y,z\}$, referred to as the Oseen tensor, and given by
\begin{equation}
	\G_{\alpha\beta}^{\mathrm{O}} (\R,\R_0) = \frac{1}{8\pi\eta} 
	\left( \frac{\delta_{\alpha\beta}}{s} + \frac{s_\alpha s_\beta}{s^3} \right) \, , 
	\label{oseenTensor}
\end{equation}
where summation over repeated indices is assumed following Einstein's convention.
Moreover $\vect{s} := \R-\R_0$ and $s := |\vect{s}|$. The flow due to a point force, called a Stokeslet, decays with the distance like $1/s$. 

The solution of the forced Stokes equations in the presence of an infinitely extended hard wall can conveniently be determined using the image solution technique~\cite{blake71}, and contains Stokeslets and higher-order flow singularities -- force dipoles and source dipoles. 
The corresponding Green's function satisfying the no-slip boundary conditions at the wall is given in term of the Blake tensor and can be presented as a sum of four contributions~\cite{kim13,blake71}
\begin{equation}
\boldsymbol{\mathcal{G}} (\R) = \boldsymbol{\mathcal{G}}^{\mathrm{O}} (\vect{s}) - \boldsymbol{\mathcal{G}}^{\mathrm{O}} (\vect{R}) + 2z_0^2\boldsymbol{\mathcal{G}}^{\mathrm{D}} (\vect{R}) - 2z_0\boldsymbol{\mathcal{G}}^\mathrm{SD} (\vect{R}) \, ,
\end{equation}
wherein $\vect{r}_0 = (0,0,z_0)$ is the point force position,  $\vect{R}:=\R - \overline{\R_0}$ with $\overline{\R_0} = (0,0,-z_0)$ is the position of the Stokeslet image with respect to the wall. 
Moreover, $r:= |\R|$ and $R:=|\vect{R}|$.
Here $\boldsymbol{\mathcal{G}}^{\mathrm{D}}$ is the force dipole given by
\begin{equation}
	\G^{\mathrm{D}}_{\alpha\beta} (\vect{R}) = \frac{(1-2\delta_{\beta z})}{8\pi\eta} 
			    \left( \frac{\delta_{\alpha\beta}}{R^3} - \frac{3R_\alpha R_\beta}{R^5} \right) \, , 
\end{equation}
and $\boldsymbol{\mathcal{G}}^\mathrm{SD}$ denotes the source dipole given by
\begin{align}
\G^\mathrm{SD}_{\alpha\beta} (\vect{R}) &= \frac{(1-2\delta_{\beta z})}{8\pi\eta}
		    \bigg(  
		    \frac{\delta_{\alpha\beta} R_z}{R^3} - \frac{\delta_{\alpha z} R_\beta}{R^3} \notag \\
		    &+ \frac{\delta_{\beta z} R_\alpha}{R^3}
		    -\frac{3R_\alpha R_\beta R_z}{R^5}
		    \bigg) \, .
\end{align}

\begin{figure}
	\includegraphics[scale=0.65]{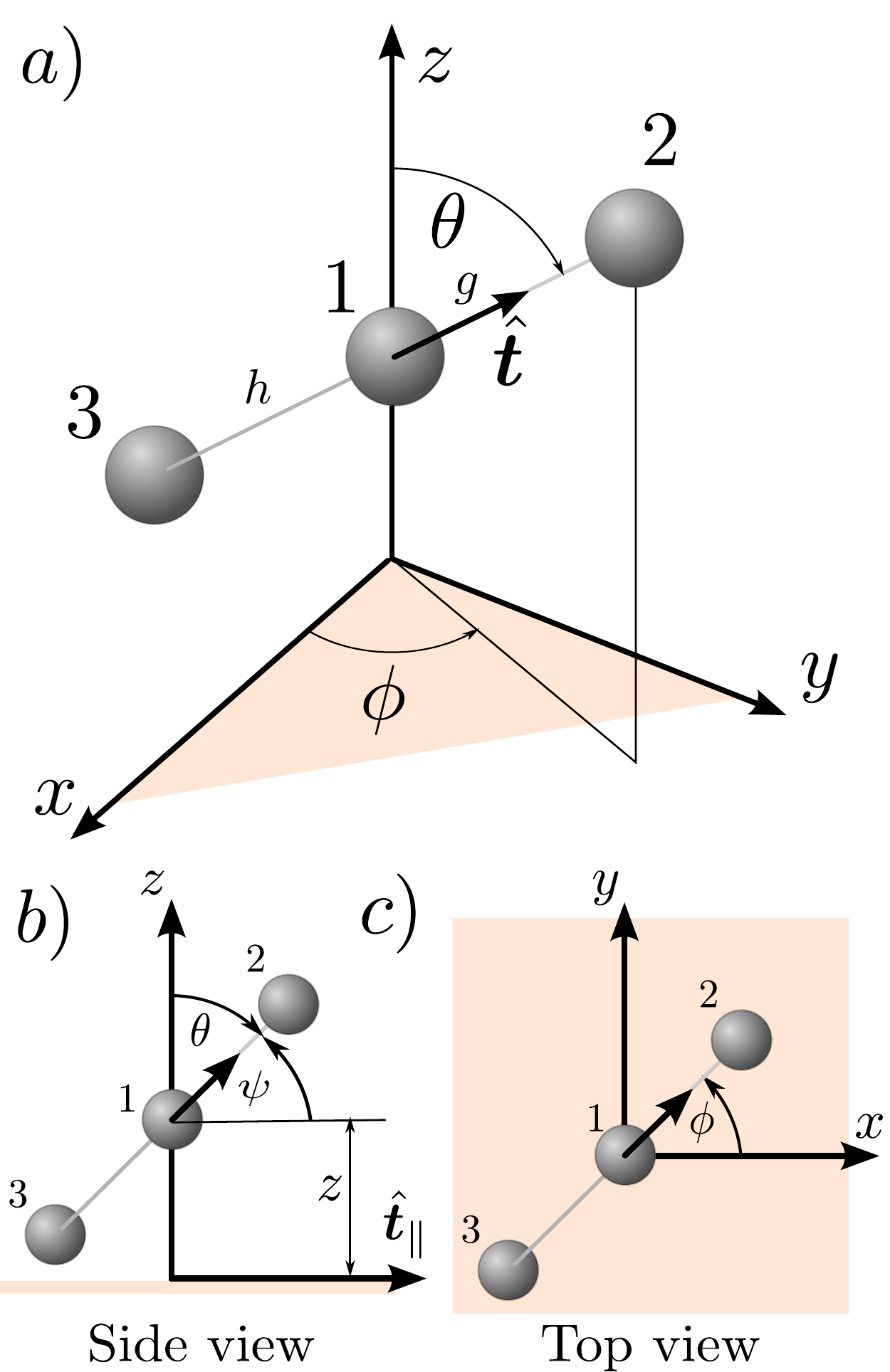}
	\caption{$a)$ The frame of reference associated with a neutral three-linked sphere low-Reynolds-number microswimmer, relative to the laboratory frame.
	The swimmer is oriented along the unit vector~$\vect{\hat{t}}$ defined by the azimuthal angle~$\phi$ and polar angle~$\theta$.
	The spheres are connected to each other by dragless rods where the instantaneous distances between the spheres 2 and 3 relative to the sphere 1 are denoted $g$ and $h$, respectively.
	{The side and top views are shown in the subfigures $b)$ and $c)$, respectively, where~$\hat{\vect{t}}_\parallel$ stands for the projection of orientation vector~$\hat{\vect{t}}$ on the plane $z=0$.
	Here $\psi:=\theta-\pi/2$.}}
	\label{illustration}
\end{figure}

The translational and rotational motion of the particles is related to the forces $\vect{F}$ and torques $\vect{L}$ acting upon them via the hydrodynamic mobility tensor. 
In the presence of a background flow with velocity $\bv_0$ and vorticity $2\bm{\omega}_0$, this relation takes the form 
\begin{equation}
	    \begin{pmatrix}
	    \BV-\bv_0 \\
	    \bm{\Omega} - \bm{\omega}_0 \\
	    \end{pmatrix} =  
	    \begin{pmatrix}
	    \mi^{tt} & \mi^{tr} \\
	    \mi^{rt} & \mi^{rr}  \\
	    \end{pmatrix}
	    \begin{pmatrix}
	    \BF  \\
	    \BL \\
	    \end{pmatrix} \, . \label{mobilityDef}
\end{equation}
The indices indicate the translational ($tt$), rotational ($rr$), and translation-rotation coupling ($tr$, $rt$) parts of the mobility tensor.
The mobility tensor contains contributions relative to a single particle (self mobilities), in addition to contributions due to interactions between the particles (hereafter approximated by pair mobilities).
Owing to the linearity of the Stokes equations and the reciprocal theorem, the hydrodynamic mobility tensor is always symmetric, and positive definite~\cite{swan07,su17,fiore17}.

\subsection{Swimmer model}

In low-Reynolds-number hydrodynamics, swimming objects have to undergo non-reciprocal motion in order to achieve propulsion.
In the present work, we use a simple model swimmer, originally proposed by Najafi and Golestanian~\cite{najafi04}, which is made of three aligned spheres. The spheres are connected by rod-like elements of negligible hydrodynamic effects in order to ensure their alignment. This system is capable of swimming forward when the mutual distances between the spheres are varied periodically in such a way that the time-reversal symmetry is broken (see Fig.~\ref{illustration} for an illustration of the linear swimmer model).
{In the present article, we focus our attention on the behavior of a neutral swimmer for which the three spheres have equal size.
The behavior of a general three-sphere microswimmer with different sphere radii to discriminate between pushers and pullers will be reported elsewhere~\cite{daddi18jpcm}.}

\subsubsection{Mathematical formulation}

Assuming that the fluid surrounding the swimmer is at rest, the translational velocity of each sphere relative to the laboratory (LAB) frame of reference is related to the internal forces $\F_\lambda$ and torques $\vect{L}_\lambda$ via the hydrodynamic mobility tensor as (c.f.~Eq.~\eqref{mobilityDef})
\begin{equation}
	\vect{V}_\gamma = \frac{\Intd \R_\gamma}{\Intd t} = \sum_{\lambda=1}^{3}
	\left( \mi_{\gamma\lambda}^{tt} \cdot \F_\lambda + \mi_{\gamma\lambda}^{tr} \cdot \vect{L}_\lambda   \right) \, , \label{linearEvolution} \\
\end{equation}
for $\gamma \in \{1,2,3\}$.
{These internal forces and torques can be actuated, e.g., by imaginary motors embedded between the spheres along the swimmer axis.}
Analogously, the angular velocity of each sphere with respect to the LAB frame is
\begin{equation}
		\vect{\Omega}_\gamma = \sum_{\lambda=1}^{3}
			\left( \mi_{\gamma\lambda}^{rt} \cdot \F_\lambda + \mi_{\gamma\lambda}^{rr} \cdot \vect{L}_\lambda   \right) \, . \label{angularEvolution}
\end{equation}
We note that $\mi_{\gamma\lambda}^{tr} = \mi_{\lambda\gamma}^{rt}$ as required by the symmetry of the mobility tensor.

Since the swimmer has to undergo autonomous motion, its body has to be both force-free and torque-free in total.
Accordingly,		
\begin{equation}
	\sum_{\lambda=1}^{3} \F_\lambda = 0 \, , \qquad
	\sum_{\lambda=1}^{3} \big( \left(\R_\lambda-\vect{r}_\mathrm{R} \right) \times \F_\lambda 
	    +    \vect{L}_\lambda \big)  = 0 \, , \label{force-free-torque-free}
\end{equation}
where $\times$ denotes the cross product. The moments of the internal forces can be taken with respect to any reference point~$\vect{r}_\mathrm{R}$, that we chose here for convenience as the position of the central sphere.

We now assume that the instantaneous relative distance vectors between the spheres are prescribed at each time as
\begin{subequations}\label{hg_Eq}
	\begin{align}
		\R_1-\R_3 = h(t) \, \vect{\hat{t}} \, ,  \\
		\R_2-\R_1 = g(t) \, \vect{\hat{t}} \, , 
	\end{align} 
\end{subequations}
where $\vect{\hat{t}}$ is the unit vector pointing along the swimming direction such that $\vect{\hat{t}} = \sin\theta \cos\phi \, \vect{\hat{e}}_x + \sin\theta\sin\phi \, \vect{\hat{e}}_y + \cos\theta \, \vect{\hat{e}}_z$ (c.f.\@ Fig.~\ref{illustration}).
Here $\phi$ and $\theta$ stand for the azimuthal and polar angles, respectively, in the spherical coordinate system associated with the swimmer.
We further define the unit vectors $\vect{\hat{\theta}} = \cos\phi\cos\theta \, \vect{\hat{e}}_x + \sin\phi\cos\theta \, \vect{\hat{e}}_y - \sin\theta \, \vect{\hat{e}}_z$ and $\vect{\hat{\phi}} = -\sin\phi \, \vect{\hat{e}}_x + \cos \phi \, \vect{\hat{e}}_y$.
We note that the set of vectors $(\vect{\hat{t}}, \vect{\hat{\theta}}, \vect{\hat{\phi}} )$ forms a direct orthonormal basis satisfying the relation $\vect{\hat{\theta}} \times \vect{\hat{\phi}} = \vect{\hat{t}}$.
Throughout this work, we assume that the lengths of the rods change periodically in time relative to a mean value $L$, 
\begin{subequations}\label{def_omega}
	\begin{align}
		g(t) &= L + u_{10} \cos(\omega t) \, , \\
		h(t) &= L + u_{20} \cos (\omega t + \delta) \, ,
	\end{align} 
\end{subequations}
where $\omega$ is the frequency of motion, $\delta \in [0,2\pi)$ is the phase shift, and $u_{10}$ and $u_{20}$ are the amplitudes of the length change such that $|u_{10}| \ll L$ and $|u_{20}| \ll L$.
For $\delta \notin\{ 0, \pi\}$ and non-vanishing $u_{10}$ and $u_{20}$, this constitutes a non-reciprocal motion, which -- as noted before -- is needed for self-propulsion at low Reynolds numbers.

By combining Eqs.~\eqref{linearEvolution}, providing the instantaneous velocities of the spheres with Eq.~\eqref{hg_Eq}, we readily obtain
\begin{subequations}
	\begin{align}
		\sum_{\lambda=1}^{3} 
				\left( \vect{G}^{tt}_\lambda \cdot \F_\lambda +  \vect{G}^{tr}_\lambda \cdot \vect{L}_\lambda   \right) 
				&= \dot{g} \, \vect{\hat{t}} + g \, \frac{\Intd \vect{\hat{t}}}{\Intd t} \, , \label{linearEvolution_21} \\
		\sum_{\lambda=1}^{3} 
				\left( \vect{H}_\lambda^{tt} \cdot \F_\lambda + \vect{H}_\lambda^{tr} \cdot \vect{L}_\lambda   \right) 
				&= \dot{h} \, \vect{\hat{t}} + h \, \frac{\Intd \vect{\hat{t}}}{\Intd t} \, , \label{linearEvolution_13} 
	\end{align}\label{linearEvolution_123} 
\end{subequations}
where, for convenience, we have defined the tensors
\begin{subequations}
	\begin{align}
		\vect{G}^{\alpha\beta}_\lambda &:=  \mi_{2\lambda}^{\alpha\beta} - \mi_{1\lambda}^{\alpha\beta} \, , \\
		\vect{H}^{\alpha\beta}_\lambda &:=  \mi_{1\lambda}^{\alpha\beta} - \mi_{3\lambda}^{\alpha\beta} \, , 
	\end{align}
\end{subequations}
with $\alpha \beta \in \{ tt,tr, rt, rr \}$.
The time derivative of the unit orientation vector $ \vect{\hat{t}}$ relative to the LAB frame is 
\begin{equation}
		\frac{\Intd \vect{\hat{t}}}{\Intd t} = \dot{\theta} \, \vect{\hat{\theta}}
		+ \dot{\phi} \, \sin\theta \,  \vect{\hat{\phi}} \, . \label{t_diff}
\end{equation}

In order to model the circular trajectories observed is swimming bacteria near surfaces, we further assume that the spheres can freely rotate around the swimming axis at rotation rates $\dot{\varphi}_\gamma$.
The frame of reference associated with the swimmer can be obtained by Euler transformations~\cite{fossen11}, consisting of three successive rotations.
Accordingly, the Euler angles~ $\phi$, $\theta$ and $\varphi_\gamma$ represent the precession, nutation, and intrinsic rotation along the swimming axis, respectively.
The angular velocity vector of a sphere~$\gamma$ relative to the LAB frame reads
\begin{equation}
	\vect{\Omega}_\gamma = -\dot{\phi}\sin\theta \, \vect{\hat{\theta}} + \dot{\theta} \, \vect{\hat{\phi}} + (\dot{\phi} \cos\theta + \dot{\varphi}_\gamma) \, \vect{\hat{t}}  \, .
	\label{angularVelocity}
\end{equation}

The dynamics of the swimmer are fully characterized by the instantaneous velocity of the central sphere in addition to the rotation rates $\dot{\theta}$ and $\dot{\phi}$.
For their calculation, we require the knowledge of the internal forces and torques acting between the spheres.

By projecting Eqs.~\eqref{linearEvolution_123} onto the spherical coordinate basis vectors and eliminating the rotation rates $\dot{\theta}$ and $\dot{\phi}$, four scalar equations are obtained. 
The force- and torque-free conditions stated by Eq.~\eqref{force-free-torque-free} provide us with six additional equations. 
Moreover, the projection of the angular velocities~\eqref{angularVelocity} along the $\vect{\hat{\theta}} $ and $\vect{\hat{\phi}}$ directions yields
\begin{subequations}\label{projOmegaThetaPhi}
	\begin{align}
		\vect{\Omega}_\gamma \cdot  \vect{\hat{\theta}} &= -\dot{\phi} \sin\theta  \, , \label{projOmegaTheta} \\
		\vect{\Omega}_\gamma \cdot  \vect{\hat{\phi}} &= \dot{\theta} \, , \label{projOmegaPhi} 
	\end{align}
\end{subequations}
for $\gamma \in \{1,2,3\}$, providing six further equations.
For a closure of the system of equations, we prescribe the relative angular velocities between the adjacent spheres as
\begin{subequations}
	\begin{align}
		\left(\vect{\Omega}_1 - \vect{\Omega}_3 \right) \cdot \vect{\hat{t}} = \dot{\varphi}_1-\dot{\varphi}_3 &=: \omega_{13} \, , \label{differenceAngVel_13} \\
		\left(\vect{\Omega}_2 - \vect{\Omega}_1 \right) \cdot \vect{\hat{t}} = \dot{\varphi}_2-\dot{\varphi}_1 &=: \omega_{21} \, . \label{differenceAngVel_21} 
	\end{align}\label{differenceAngVel_123}
\end{subequations}

The determination of the internal forces and torques acting on each sphere is readily achievable by solving the resulting linear system composed of 18 independent equations given by \eqref{force-free-torque-free}, \eqref{linearEvolution_123}, \eqref{projOmegaThetaPhi} and \eqref{differenceAngVel_123}, using the standard substitution method.
In the remainder of this paper, all the lengths will be scaled by the mean length of the arms~$L$ {and the times by the inverse frequency~$\omega^{-1}$.}
Finally, the swimming velocity can be calculated as 
\begin{equation}
	\vect{V} := \vect{V}_1 = \sum_{\lambda=1}^{3}
		\left( \mi_{1\lambda}^{tt} \cdot \F_\lambda + \mi_{1\lambda}^{tr} \cdot \vect{L}_\lambda   \right) \, . \label{velocityMean}
\end{equation}
and the rotation rates as
\begin{align}
	\dot{\theta} &= \frac{1}{h} \sum_{\lambda=1}^{3} \left( \vect{H}_\lambda^{tt} \cdot \F_\lambda + \vect{H}_\lambda^{tr} \cdot \vect{L}_\lambda   \right) \cdot \vect{\hat{\theta}} \, , \label{thetaDot} \\
		\dot{\phi}  &= \frac{1}{h \, \sin\theta}  \sum_{\lambda=1}^{3} \left( \vect{H}_\lambda^{tt} \cdot \F_\lambda + \vect{H}_\lambda^{tr} \cdot \vect{L}_\lambda   \right) \cdot \vect{\hat{\phi}} \, . \label{phiDot}
\end{align}
The swimming trajectories can thus be determined by integrating Eqs.~\eqref{velocityMean} through \eqref{phiDot} for a given set of initial conditions $(\R_0,\theta_0,\phi_0)$.

\subsubsection{{Swimming in an unbounded domain}}

{In an unbounded fluid domain,} i.e., in the absence of the wall, the swimmer undergoes purely translational motion along its swimming axis without changing its orientation.
In order to proceed analytically, we assume that the radius of the spheres~$a$ is much smaller than the arm lengths.
The internal forces acting on the spheres averaged over one swimming period are 
\begin{equation}
	\F_1 = \frac{a^2}{4} \left( 5 +\frac{11}{2} \, a \right) \pi \eta K \, \vect{\hat{t}} \, , \qquad 
	\F_2 = \F_3 = -\frac{\F_1}{2} \, , 
\end{equation}
wherein
\begin{equation}
	K:= \langle g\dot{h}-h\dot{g} \rangle = -u_{10} u_{20} \sin \delta \, , 
\end{equation}
and $\langle \cdot \rangle$ denotes the time-averaging operator over one complete swimming cycle, defined by
\begin{equation}
	\langle \cdot \rangle := \frac{1}{2\pi} \int_0^{2\pi} (\cdot) \, \Intd t \, .
\end{equation}
Clearly, no net swimming motion is achieved if $\delta =0$ or $\pi$. 
Moreover, the swimming speed is maximal when $\delta = \pi/2$, a value we consider in the subsequent analysis.
The internal torques exerted on the rotating spheres read
\begin{subequations}
	\begin{align}
		\vect{L}_1 &= \frac{8\pi}{3} \, a^3 \left( \omega_{13}-\omega_{21} \right) \vect{\hat{t}} \, , \\
		\vect{L}_2 &= \frac{8\pi}{3} \, a^3 \left( 2\omega_{21}+\omega_{13} \right) \vect{\hat{t}} \, , \\
		\vect{L}_3 &= -\frac{8\pi}{3} \, a^3 \left( \omega_{21}+2\omega_{13} \right) \vect{\hat{t}} \, .
	\end{align}
\end{subequations}

By making use of Eq.~\eqref{velocityMean} and averaging over a swimming cycle, the translational velocity up to the second order in $a$ reads
\begin{equation}
	\vect{V}_1 = V_0 \, \vect{\hat{t}} \, , \qquad V_0 := -\frac{a}{24} \left(7 + 5a\right) K \, , 	\label{velocityBulk}
\end{equation}
while $\dot{\theta} = 0$ and $\dot{\phi} = 0$ so that the swimmer's orientation remains constant.
{Evidently}, the averaged swimming speed is a function of just the swimmer's properties and does not depend on the fluid viscosity~\cite{najafi04}.
The fluid viscosity would nevertheless have to be accounted for to calculate the power needed to perform the prescribed motions of the three spheres. 
In the following, we will address the swimming behavior near a hard wall and investigate the possible scenarios of motion.

\begin{figure}
\begin{center}
\includegraphics[scale=1]{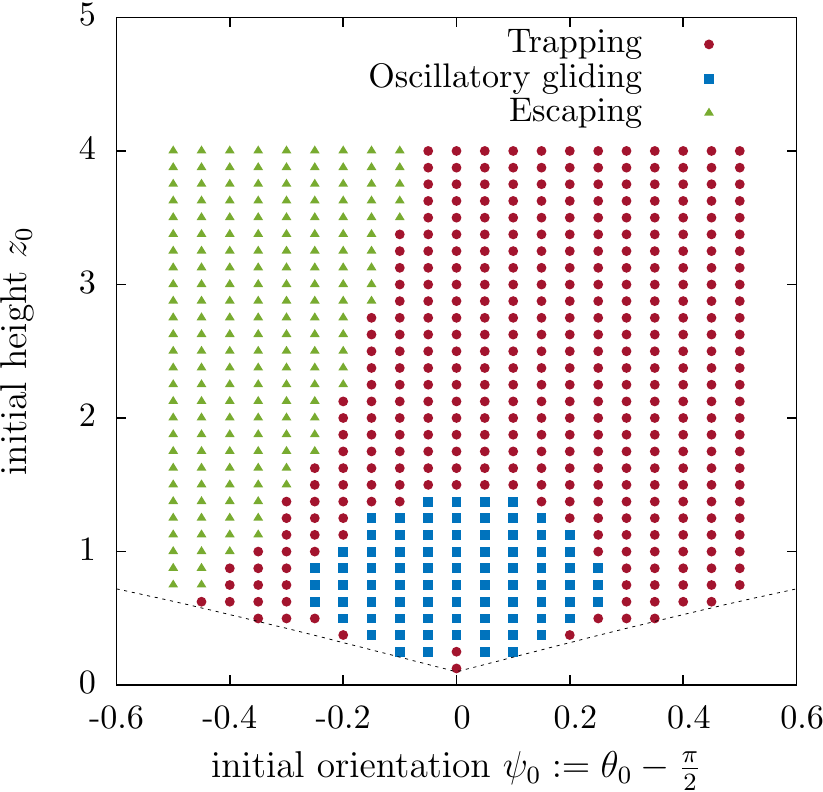}
\end{center}
\caption{(Color online) State diagram illustrating the possible swimming scenarios in the presence of a hard wall for the 2D motion, i.e. for $\omega_{21}=\omega_{13}=0.$
The dashed line corresponds to impermissible situations in which one of the spheres is in contact with the wall.
Here $a=u_{10}=u_{20} = 1/10$.
}
\label{Phase-Diagram-Hard-Wall}
\end{figure}

\section{Swimming near a wall}\label{section-state-diagram}

\subsection{State diagram}

\begin{figure}
\begin{center}
\includegraphics[scale=1]{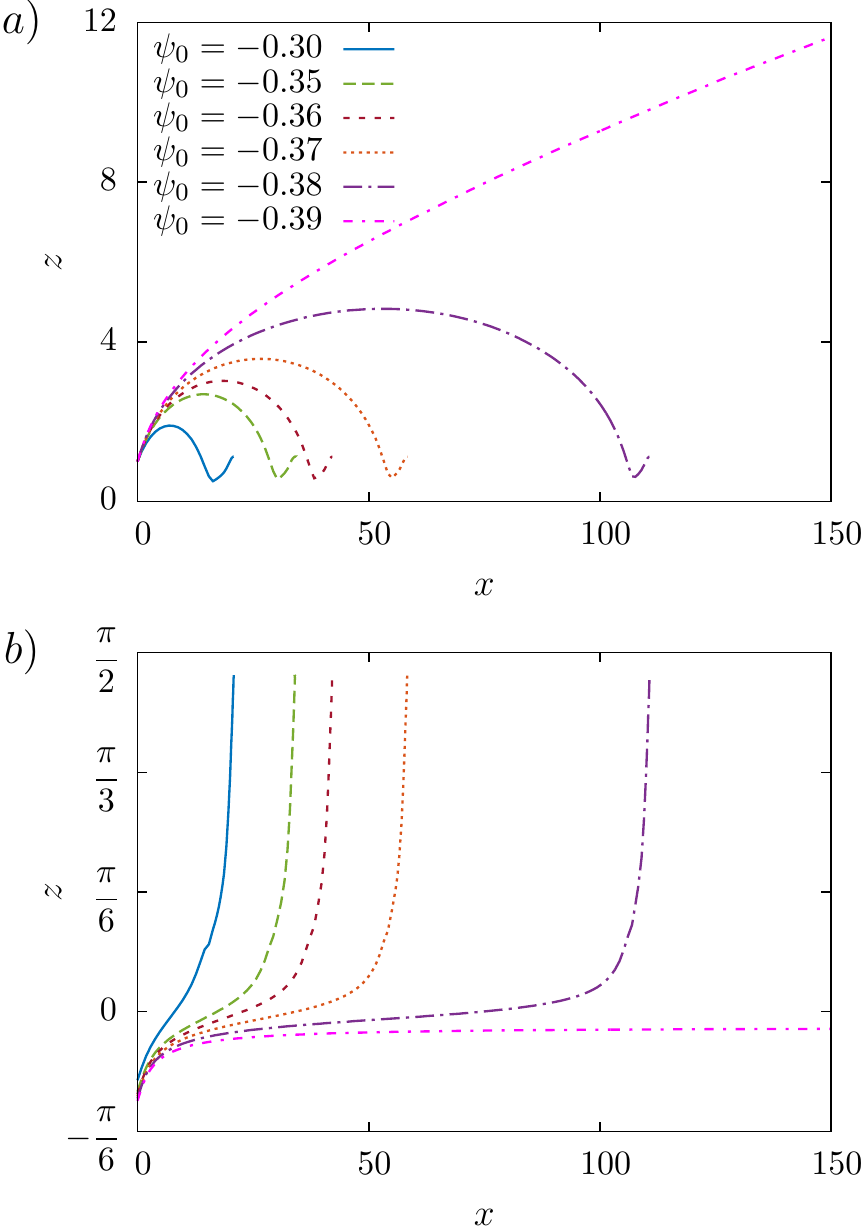}
\end{center}
\caption{(Color online) Transition from the trapping to the escaping states upon variation of the initial inclination angle~$\psi_0$ while keeping the initial distance from the wall constant at $z_0 = 1$.
$a)$ shows the averaged swimming trajectories for the 2D motion in the plane $(x,z)$ and $b)$ the inclination angle $\psi$ as a function of~$x$.}
\label{Escape}
\end{figure}

\begin{figure}
\begin{center}
\includegraphics[scale=1]{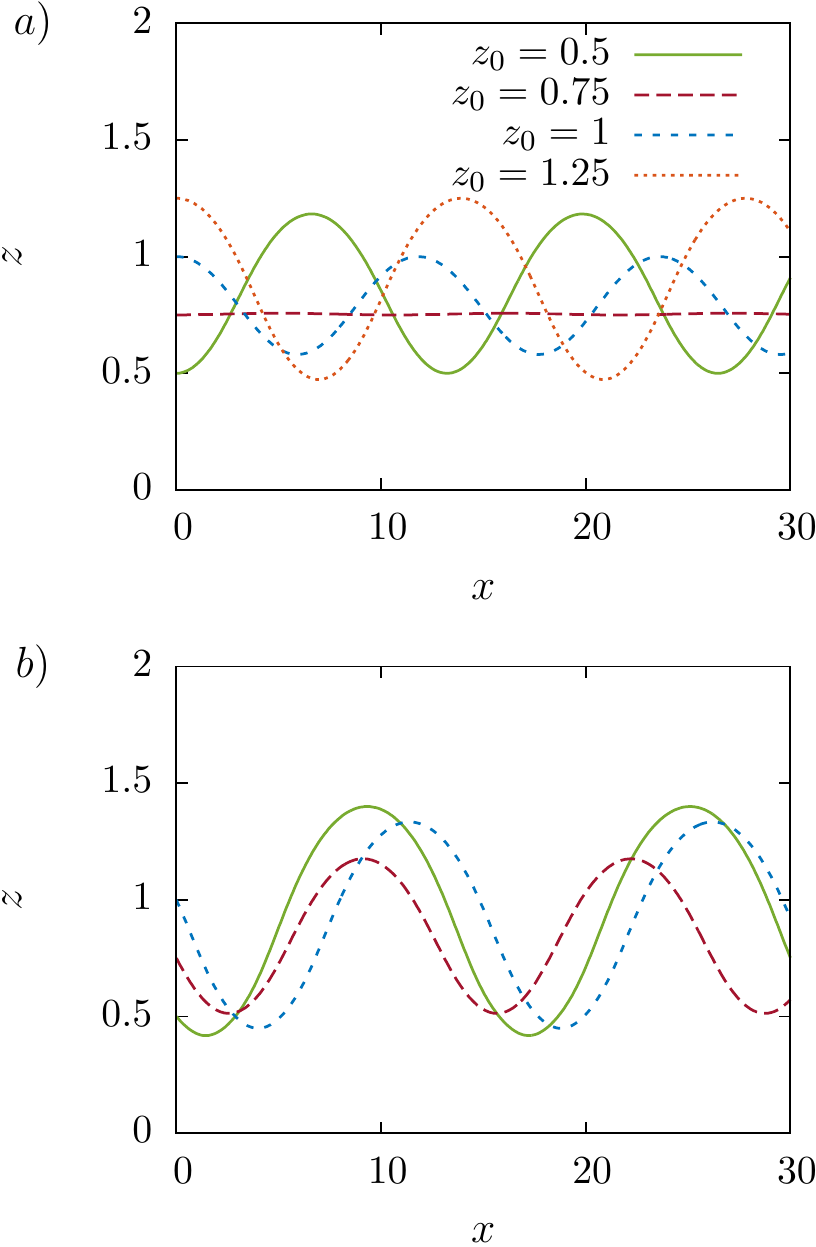}
\end{center}
\caption{(Color online) Typical swimming trajectories in the oscillatory gliding state for different initial distances from the wall {where $a)$ $\psi_0 = 0$ and $b)$ $\psi_0=0.2$.
For $z_0=1.25$ and $\psi_0=0.2$, the swimmer is trapped by the wall and thus the trajectory has not been shown here.
}
The swimmer inclination angle shows a similar oscillatory behavior around a mean angle $\psi=0$.}
\label{Oscillations}
\end{figure}

We now consider the swimming kinematics in the vicinity of a hard wall and examine in details the resulting swimming trajectories.
For that aim, we solve numerically the linear system of equations described in the previous section to determine the internal forces and torques acting between the spheres. 
The time-dependent position and orientation of the swimmer are then calculated by numerically integrating  Eqs.~\eqref{velocityMean} through \eqref{phiDot} using a fourth-order Runge-Kutta scheme with {adaptive time stepping}~\cite{press92}.  
For the particle hydrodynamic mobility functions, we employ the values obtained using the multipole method for Stokes flows~\cite{cichocki00, kkedzierski10}.
This method is widely used and has the advantage of providing precise and accurate predictions of the self-mobilities, which are reasonable even at distances very close to the wall. 
The time-averaged positions and inclinations are determined numerically using the standard trapezoidal integration method.
As the vertical position of one of the spheres gets closer to the wall such that $z \sim a$, an additional soft repulsive force $F_z = \kappa (z-a)^{-n}$ is introduced, where $\kappa = 10^{-5} \eta |K|$ and $n=2$ are taken as typical values.
We have checked that changing these values within moderate ranges results in qualitatively similar outcomes.
Moreover, we take $a=u_{10}=u_{20}=1/10$.

We begin with the relatively simple situation in which the spheres do not rotate around the swimming axis, so we take $\omega_{21} = \omega_{13} = 0$.
In this particular case, the problem becomes two dimensional as the swimmer is constrained to move in the plane defined by its initial azimuthal orientation $\phi_0$.
Without loss of generality, we take $\phi_0=0$ for which the swimmer moves in the $(x,z)$ plane.

\begin{figure*}
\begin{center}
\includegraphics[scale=1]{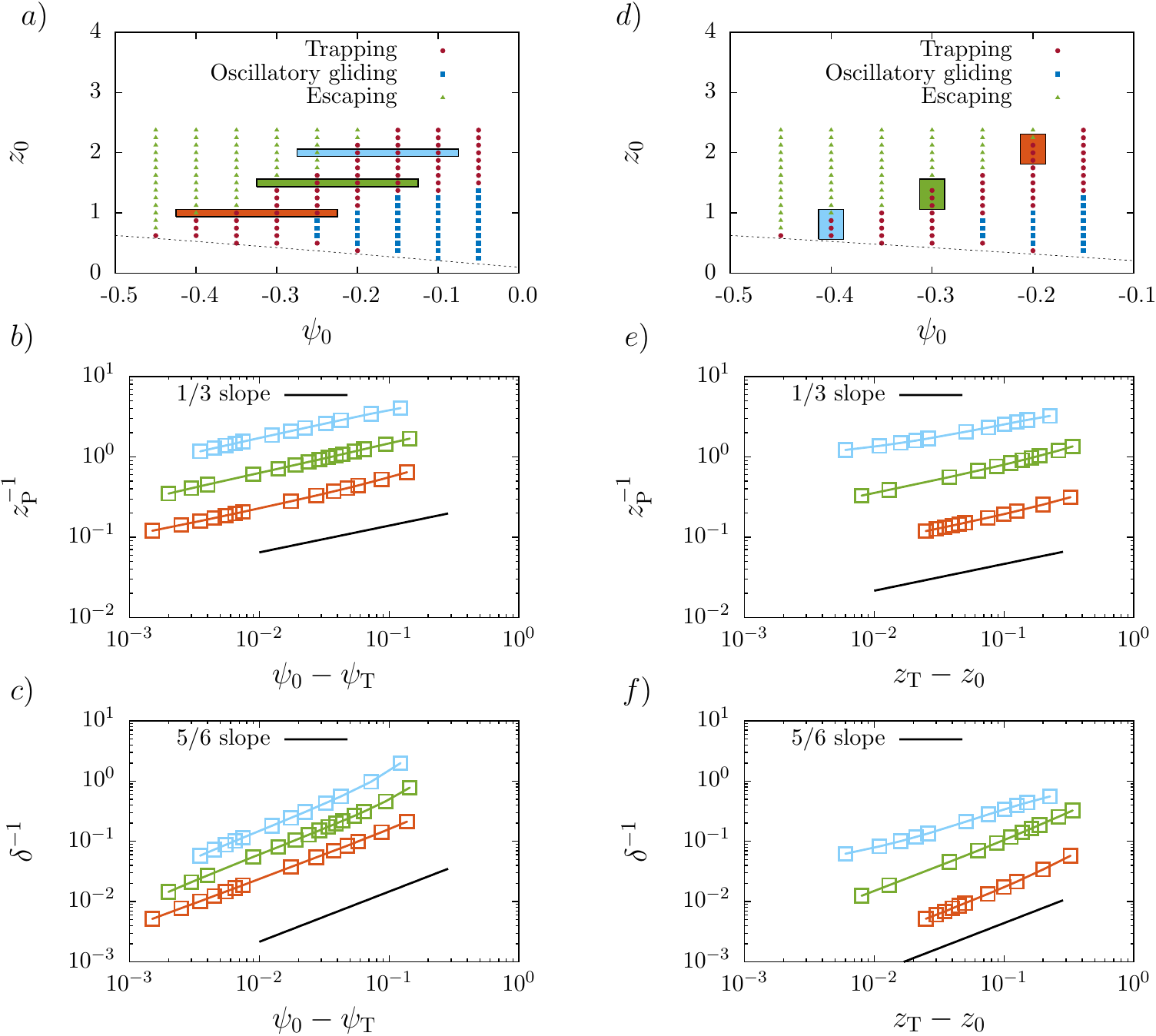}
\end{center}
\caption{(Color online) Log-log plots of order parameters $\zP^{-1}$ and $\delta^{-1}$ at the transition point between the trapping and escaping states in the 2D case for $\omega_{13}=\omega_{21} = 0$, as obtained from the numerical simulations.
{Here $\zT$ and $\psiT$ denote respectively the swimmer height and inclination at the transition point between the trapping and escaping states.
For the sake of readability, the curves  associated with the green and blue paths are shifted on the vertical scale by factors of 3 and 9, respectively.}
The solid lines are a guide for the eye.
}
\label{Fig-Esc_Both_Shifted}
\end{figure*}

In Fig.~\ref{Phase-Diagram-Hard-Wall}, we show the swimming state diagram constructed in the $(z_0,\psi_0)$ space, where $\psi := \theta-\pi/2$ defines the angle relative to the horizontal direction.
Hence, the swimmer is initially pointing towards (away from) the wall for $\psi_0>0$ ($\psi_0<0$).
We observe that three different possible scenarios of motion emerge depending upon the initial distance from the wall and orientation. 
The swimmer may be trapped by the wall, totally escape from the wall, or undergo a nontrivial oscillatory gliding motion.
In the trapping state (shown as red circles in Fig.~\ref{Phase-Diagram-Hard-Wall}), the swimmer moves towards the wall following a parabolic-like trajectory to progressively align perpendicular to the wall as $\psi \to \pi/2$.
In the final stage, the swimmer reaches a stable state and hovers at a constant height above the wall.
This behavior occurs for large initial inclinations when $\psi_0 > 0.3$ and that regardless of the initial distance that separates the swimmer from the wall.
However, trapping can also take place for $\psi_0 \sim 0$ if the swimmer is initially located far enough from the wall, at distances larger than $z_0 = 1.5$.
Notably, the swimmer is trapped by the wall if it is released from distances $z_0<0.25$ with a vanishing initial inclination $\psi_0=0$.

The escaping state (green triangles in Fig.~\ref{Phase-Diagram-Hard-Wall}) is observed if the swimmer is directed away from the wall with $\psi_0<-0.5$.
In this state, the swimmer moves straight away from the wall beyond a certain height at which the wall-induced hydrodynamic interactions die away completely.	
In the oscillatory gliding state (blue rectangles in Fig.~\ref{Phase-Diagram-Hard-Wall}), the swimmer undergoes a sinusoidal-like motion around a mean height above the wall.
This state occurs in a bounded region of initial states when $z_0 \sim 1$ and $\psi_0 \sim 0$.


In Fig.~\ref{Escape}, we show the transition from the trapping to the escaping states upon variation of the initial inclination for a swimmer initially positioned a distance $z_0=1$ above the wall.
{For initial inclinations $\psi_0 > -0.39$,} the swimmer moves along a curved path following a projectile-like trajectory before ending up hovering at a steady height $z \simeq 1.12$ above the wall.
Accordingly, the swimmer velocity normal to the wall vanishes and the inclination angle approaches the steady value corresponding to $\psi \simeq \pi/2$.
Indeed, this final state is stable and is found to be independent of the initial orientation of the swimmer with respect to the wall.
For $\psi_0 = -0.39$, the swimmer manages to escape from the attraction of the wall and moves along a straight line maintaining a constant orientation, i.e.\@ just as it would be the case in an unbounded fluid.

Fig.~\ref{Oscillations} illustrates the swimming trajectories in the oscillatory gliding state  {for $a)$ $\psi_0 = 0$ and $b)$ $\psi_0=0.2$ and various initial heights} ranging from $z_0=0.5$ to 1.25. 
We observe that the amplitude of oscillations is strongly dependent on $z_0$, and eventually vanishes {for $\psi_0=0$} and $z_0 \simeq 0.75$ giving rise to a steady sliding motion at a constant velocity.
The mean inclination angle over one oscillation period amounts to zero and thus the swimmer undergoes motion at a constant mean height above the wall.
We further note that the frequency of oscillations has nothing to do with $\omega$ which is several orders of magnitude larger.

{For future reference, we denote by $\mu$ the magnitude of the scaled swimming velocity parallel to the wall averaged over one oscillation period, $\mu:= \overline{V_\parallel}/V_0$ where $V_\parallel:= \left(V_x^2+V_y^2\right)^{1/2}$ and $V_0$ is the magnitude of the bulk swimming velocity given by Eq.~\eqref{velocityBulk}.}

\subsection{Transition between states}

We now investigate the swimming behavior more quantitatively and analyze the evolution of relevant order parameters around the transition points between the states.

\subsubsection{Transition between the trapping and escaping states}

In order to probe the transition between the trapping and escaping states, we define an order parameter $\zP^{-1}$ as the inverse of the peak height achieved by the swimmer before it is trapped by the wall (c.f.~Fig.~\ref{Escape}~$a)$).
Additionally, we define a second order parameter $\delta^{-1}$ as the inverse of the distance along the $x$ direction at which the peak height occurs.
Clearly, both $\zP^{-1}$ and $\delta^{-1}$ amount to zero for the escaping state, and thus can serve as relevant order parameters to characterize the transition between the trapping and escaping states.

In Fig.~\ref{Fig-Esc_Both_Shifted}, we present the evolution of the order parameters $\zP^{-1}$ and $\delta^{-1}$ around the transition point between the trapping and escaping states along three different horizontal (subfigures $a), b),$ and $c)$) and vertical (subfigures $d), e),$ and $f)$) paths in the state diagram presented in Fig.~\ref{Phase-Diagram-Hard-Wall}.
We observe that the {inverse peak height $\zP^{-1}$} exhibits a scaling behavior around the transition points with an exponent of 1/3.
Similar behavior is displayed by {the inverse peak position} around the transition points with a scaling exponent of 5/6.
We will show in Sec.~\ref{farfield_order} that these scaling laws can indeed be predicted theoretically by considering a simplified model based on the far-field approximation.
{It can clearly be seen that even beyond $\psi-\psi_0=0.1$ from the transition points, the scaling law is still approximatively obeyed.}
Despite its simplicity, the presented far-field model leads to a good prediction of the scaling behavior of these two order parameters around the transition points.

\subsubsection{Transition between the trapping and oscillatory-gliding states}

\begin{figure}
\begin{center}
\includegraphics[scale=1]{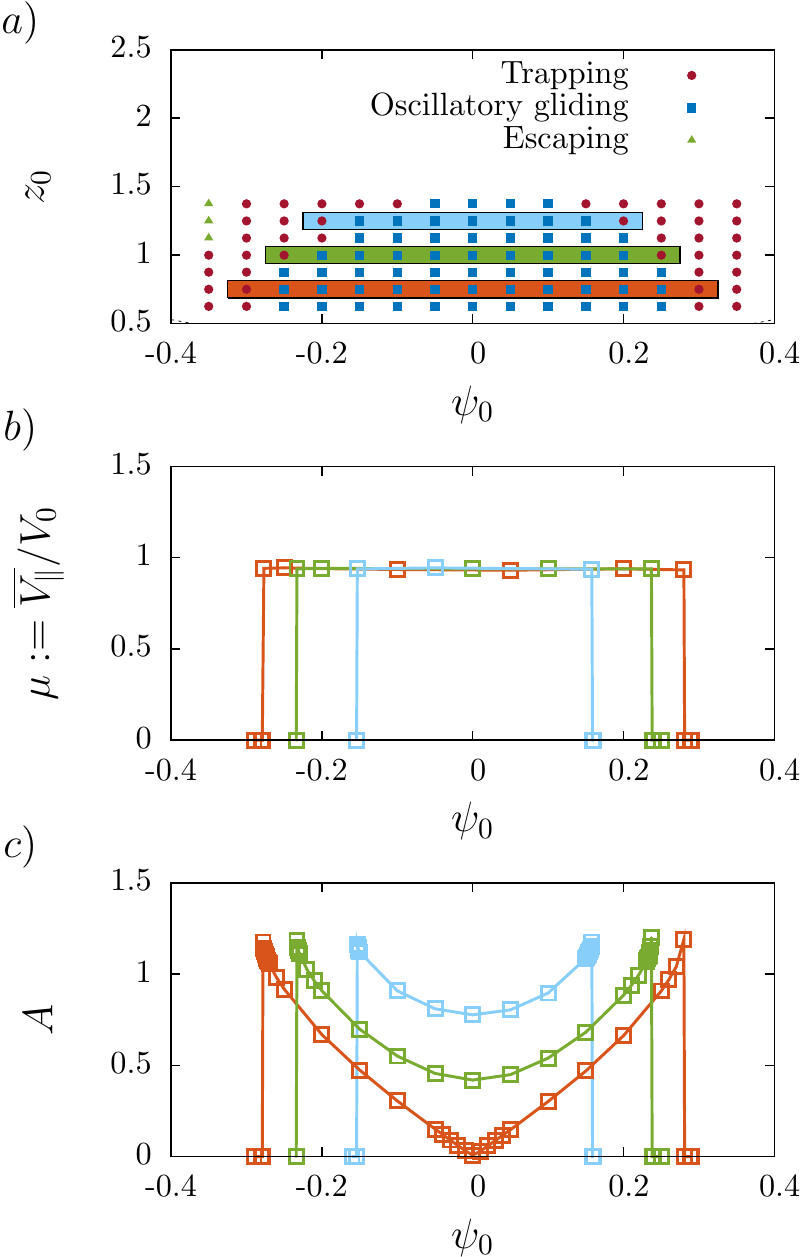}
\end{center}
\caption{(Color online) Evolution of the order parameters $a)$ $\mu$ and $b)$ $A$ versus the initial inclination angle $\psi_0$ at the transition between the trapping and oscillatory gliding states for various horizontal paths along the state diagram.}
\label{Fig-Osc_Mean}
\end{figure}

In the oscillatory-gliding state, the swimmer remains on average at the same height above the wall such that $\overline{V_z} = 0$ and translates at a constant velocity parallel to the wall.
In order to study the transition between the trapping and oscillatory-gliding states, {we utilize the scaled mean swimming  velocity parallel to the wall, averaged over one oscillation period as a relevant order parameter, $\mu = \overline{V_x}/V_0$, where again $V_0$ is the magnitude of the swimming velocity in an unbounded fluid domain.}
Additionally, we define a second order parameter $A$ as the amplitude of oscillations.

In Fig.~\ref{Fig-Osc_Mean}, we present the evolution of the order parameters $\mu$ and $A$ at the transition points between the oscillatory-gliding and trapping states along three different horizontal paths in the state diagram.
The mean swimming velocity (Fig.~\ref{Fig-Osc_Mean}~$a)$) is found to be about $5\%$ lower than the bulk velocity and is weakly dependent on the initial orientation or distance from the wall.
In the trapping state, the swimmer points toward the wall and remains at a constant height above the wall to attain a stable hovering state.
Therefore, in this situation, both of the two order parameters $\mu$ and $A$ vanish.
The transition from the oscillatory-gliding and trapping states is thus first order, characterized by a discontinuity in the relevant order parameters. 
We further remark that the amplitudes of oscillations (Fig.~\ref{Fig-Osc_Mean}~$b)$) reach a maximum value of about 1.2 around the transition points between the oscillatory-gliding and trapping states. 
Moreover, for $\psi_0 = 0$, the amplitude of oscillations is minimal and eventually vanishes for $z_0\simeq0.75$, leading to a pure gliding motion of vanishing amplitude, parallel to the wall. 
Both order parameters are found to be symmetric with respect to $\psi_0=0$, and thus $(z_0,\psi_0)$ and $(z_0,-\psi_0)$ represent identical dynamical states along these considered paths.

In the next section, we will present a far-field model for the near-wall swimming and provide theoretical arguments for the scaling behavior observed at the transition between the trapping and escaping states.

\section{Far-field model}\label{section-far-field-model}

In order to address the swimming behavior in the far-field limit, we expand the averaged translational velocity and rotation rate of the swimmer as power series in the ratio $1/z$.
We further employ the far-field expressions of the hydrodynamic mobility functions
which can adequately be expressed as power series in the ratio $a/z$.
Up to the second order in $a$, and by accounting for the leading order in $1/z$ only, the differential equations governing the averaged dynamics of the swimmer far away from the wall read
\begin{subequations}
	\begin{align}
		\frac{\Intd x}{\Intd t} &=  -aK \cos\psi \Bigg(  \frac{7}{24}+\frac{3\sin^2\psi \left( 12-\cos^2\psi \right)}{64z^3} \notag \\
		&+a  \left( \frac{5}{24} + \frac{620-453\cos^2\psi+120\cos^4\psi}{1024z^3} \right)	\Bigg)
		 \, , \label{diffXdiffT} \\
		\frac{\Intd z}{\Intd t} &=  aK \sin\psi \Bigg(  \frac{7}{24}+\frac{3\left(8-16\cos^2\psi+\cos^4\psi \right)}{64z^3}  \notag \\
		&+a \left( \frac{5}{24} + \frac{158-111\cos^2\psi+30\cos^4\psi}{256 z^3} \right) \Bigg) \, , \label{diffZdiffT} \\
		\frac{\Intd \psi}{\Intd t} &= -\frac{9aK }{512z^4} \, \cos\psi
		\bigg( 56-52\cos^2\psi+11\cos^4\psi  \notag \\
		&+ \frac{a}{2} 
		\left( 68-31\cos^2\psi + 8\cos^4\psi \right) \bigg)	\, . \label{diffThetadiffT}
	\end{align}\label{diffXZThetadiffT}
\end{subequations}

The wall-induced correction to the swimmer translational velocities decay in the far field as $z^{-3}$ whereas its angular velocity undergoes a decay as $z^{-4}$.
Therefore, the flow field induced by a neutral three-linked sphere swimmer near a wall resembles that of a microorganism whose flow field is modeled as a force quadrupole or a source dipole.

We recall that the swimming trajectories resulting from quadrupolar hydrodynamic interactions as derived from \Faxen's law for a prolate ellipsoid of aspect ratio $\gamma$ tilted an angle $\psi$ and located a distance $z$ above a rigid wall read~\cite{spagnolie12}
\begin{subequations}
	\begin{align}
		 \frac{\Intd x}{\Intd t} &= \cos\psi \left( V_0 + \cfrac{\sigma}{16 z^3} \left( 27\cos^2\psi-20 \right) \right)  \, ,  \\
		 \frac{\Intd z}{\Intd t} &= -\sin\psi \left( V_0 + \cfrac{\sigma }{4 z^3} \left( 9\cos^2\psi-2 \right) \right) \, ,   \\
			\frac{\Intd \psi}{\Intd t} &= \cfrac{3\sigma\cos\psi}{32 z^4} \left( 8(\Gamma-1)+6(\Gamma+2)\cos^2\psi -3\Gamma\cos^4\psi \right), 
	\end{align}
\end{subequations}
where $V_0$ is the propulsion velocity in a bulk fluid, i.e.\@ far away from boundaries and $\Gamma := (\gamma^2-1)/(\gamma^2+1)$ is the shape factor.
In addition, $\sigma$ is the quadrupole strength (has the dimension of velocity $\times$ length$^3$) where $\sigma>0$ for swimmers with small bodies and elongated flagella and $\sigma<0$ in the opposite situation~\cite{lauga09, mathijssen16}.
The equations governing the dynamics of a swimming microorganism near a wall, whose generated flow field is modeled as a source dipole read~\cite{spagnolie12}
\begin{subequations}
	\begin{align}
		 \frac{\Intd x}{\Intd t} &= \cos\psi \left( V_0 -\cfrac{\alpha}{4 z^3} \right)  \, ,   \\
		 \frac{\Intd z}{\Intd t} &= -\sin\psi \left(V_0 - \cfrac{\alpha}{ z^3} \right) \, ,   \\
			\frac{\Intd \psi}{\Intd t} &= -\cfrac{3\alpha\cos\psi}{16 z^4} \left( 2+3\Gamma(2-\cos^2\psi) \right)   \, , 
	\end{align}
\end{subequations}
where $\alpha$ is the source dipole strength (has the dimension of velocity $\times$ length$^3$) {such that $\alpha>0$ for ciliated swimming organisms which rely on local surface deformation to propel themselves through the fluid~\cite{lauga09},}
and $\alpha<0$ for non-ciliated microorganisms with helical flagella. 
Therefore, the effect of the wall on the dynamics of a three-linked sphere swimmer can conveniently be modeled as a superposition of a quadrupole of strength $\sigma>0$ and a source dipole of strength $\alpha<0$.

Notably, in the limit $z \to \infty$, Eqs.~\eqref{diffXdiffT} and \eqref{diffZdiffT} reduce to Eq.~\eqref{velocityBulk} providing the swimming velocity in an unbounded bulk fluid.
We further note that the asymptotic results derived in Ref.~\onlinecite{Zargar2009} have been reported with an erroneous  far field decay that we correct here.

\subsection{Approximate swimming trajectories}

For small inclination angles relative to the horizontal plane such that $\psi \ll 1$, the sine and cosine functions can be approximated using Taylor series expansions around $\psi=0$ where $\sin\psi \sim \psi$ and $\cos\psi \sim 1$.
We have checked that accounting for the term with $\psi^2$ in the series expansion of $\cos\psi$ has a negligible effect on the swimming trajectories, and thus has been discarded here for simplicity.
Further, restricting to the leading order in $a$, Eqs.~\eqref{diffXZThetadiffT} can thus be approximated as
\begin{subequations}
	\begin{align}
		\frac{\Intd x}{\Intd t} &= - \frac{7}{24} \,  aK \, , \label{eqDiffX1} \\
		\frac{\Intd z}{\Intd t} &= aK \left( \frac{7}{24}-\frac{21}{64} \frac{1}{z^3} \right) \psi \, , \label{eqDiffZ1} \\
		\frac{\Intd \psi}{\Intd t} &= -\frac{135 }{512} \frac{aK}{z^4}   \, . \label{eqDiffTheta1}
	\end{align}  \label{approximate-equations-2D}
\end{subequations}

Based on these equations, we now derive approximate swimming trajectories analytically.
By combining Eqs.~\eqref{eqDiffZ1} and \eqref{eqDiffTheta1} and eliminating the time differential $\Intd t$, the equation relating the swimmer inclination to its vertical position reads
\begin{equation}
	\psi\, \Intd \psi = -\frac{405}{56} \frac{\Intd z}{z(8z^3-9)} \, ,
\end{equation}
which can readily be solved subject to the initial condition of inclination and distance from the wall $(\psi_0,z_0)$ to obtain
\begin{equation}
	\exp \left( \frac{28}{15} \left(\psi^2-\psi_0^2 \right)\right) =  \frac{z^3}{z_0^3} \frac{8z_0^3-9}{8z^3-9}  \, .
	\label{diffTheta2}
\end{equation}

When the swimmer reaches its peak position, the inclination angle necessarily vanishes (provided that the swimmer is initially pointing away from the wall such that $\psi_0 < 0$).
Solving Eq.~\eqref{diffTheta2} for $\psi = 0$, the peak height can thus be estimated as
\begin{equation}
	\zP = \frac{z_0}{\left( H+\frac{8}{9} \left( 1-H \right) z_0^3 \right)^{1/3}} \, , 
\end{equation}
where we have defined the parameter $H \simeq 1+\beta \, \psi_0^2$ with $\beta = 28/15$.

\subsection{Order parameters}
\label{farfield_order}
\subsubsection{{Inverse peak height~$\zP^{-1}$}}

We now calculate the first order parameter~$\zP^{-1}$ governing the transition between the trapping and the escaping states, defined in the previous section as the inverse of the peak height, 
\begin{equation}
	 \zP^{-1} = \frac{1}{z_0} \left( H+\frac{8}{9} \left( 1-H \right) z_0^3 \right)^{1/3} \, .
\end{equation}
At the transition to the escaping state, the order parameter~$\zP^{-1}$ amounts to zero.
For a given initial inclination $\psi_0$, the transition height is estimated as 
\begin{equation}
	\zT = \frac{1}{2} \left( \frac{9H}{H-1} \right)^{1/3} \, .
	\label{zKritik}
\end{equation}
Similar, the inclination angle at the transition point between the trapping and the escaping states for a given initial vertical distance $z_0$ reads
\begin{equation}
	\psiT = - \frac{1}{14} \left( \frac{105}{\frac{8}{9} \, z_0^3-1} \right)^{1/2} \, .
	\label{thetaKritik}
\end{equation}

The scaling behavior of the order parameter~$\zP^{-1}$ around the transition point can readily be obtained by performing a Taylor series expansion around $\psi_0 = \psiT$ and $z_0=\zT$  to obtain
\begin{subequations}
	\begin{align}
		\zP^{-1} &= \frac{1}{z_0} \left(\frac{2}{-\psiT} \right)^{1/3} \left( \psi_0-\psiT \right)^{1/3} + \bigO \left(  \left( \psi_0-\psiT \right)^{4/3}   \right) \, ,
		\label{q-scaling-Theta} \\
		\zP^{-1} &= \frac{(3H)^{1/3}}{\zT^{4/3}} \left( \zT-z_0 \right)^{1/3} + \bigO \left(  \left( \zT-z_0 \right)^{4/3}   \right) \, .
			\label{q-scaling-Z}
	\end{align}
\end{subequations}
Therefore, the transition between the trapping and escaping states is continuous and characterized by a scaling exponent $1/3$ of the order parameter.

\subsubsection{{Inverse peak position~$\delta^{-1}$}}

We next calculate the second order parameter~$\delta^{-1}$, defined earlier as the inverse of the horizontal position $\delta$ corresponding to the occurrence of the peak, i.e.\@ $z(x=\delta)=z_\mathrm{P}$.
Combining Eqs.~\eqref{eqDiffX1} and \eqref{eqDiffTheta1} together, we obtain
\begin{equation}
 \frac{	\Intd x}{\Intd \psi} = \frac{448}{405} \, z^4  \, , \label{eqDiffX2}
\end{equation}
where the $\psi$-dependence of the variable $z$ can readily be obtained from Eq.~\eqref{diffTheta2} and is expressed as
\begin{equation}
	z= \frac{r^{1/3} z_0}{\left( 1+\frac{8}{9} (r-1)z_0^3 \right)^{1/3}} \, ,  \label{zAsFctTheta}
\end{equation}
where we have defined 
\begin{equation}
	r \simeq 1 + \beta \left( \psi^2-\psi_0^2 \right) \, .
\end{equation}

By inserting Eq.~\eqref{zAsFctTheta} into Eq.~\eqref{eqDiffX2}, making the change of variable $r = 1 - \beta \psi_0^2 v$, and noting the relation between the differentials,
\begin{equation}
	\Intd \psi = -\frac{1}{2\beta} \frac{\Intd r}{\left( \psi_0^2 + \beta^{-1} \left(r-1\right) \right)^{1/2}} \, , 
\end{equation}
the $x$-position corresponding to the occurrence of the peak follows forthwith upon integration of both sides of the resulting differential equation to obtain
\begin{equation}
	\delta = -\frac{224}{405} \, z_0^4 \psi_0 
	\int_0^1 \left( \frac{1-\beta \psi_0^2 v}{1-\frac{8}{9} \, \beta \psi_0^2 z_0^3 v} \right)^{4/3} \frac{\Intd v}{(1-v)^{1/2}} \, .
	\label{deltaEquationIntegral}
\end{equation}

Unfortunately, the latter integral cannot be solved analytically for arbitrary values of $\psi_0$ and $z_0$.
In order to overcome this difficulty, we may have recourse to approximate analytical tools.
Clearly, there are no issues coming from the factor $\left( 1-\beta \psi_0^2 v \right)^{4/3} (1-v)^{-1/2}$ since it is well behaved and integrable in the interval $[0,1]$.
However, difficulties arise from the factor $\left(1-\frac{8}{9} \, \beta \psi_0^2 z_0^3 v\right)^{-4/3}$, in which, for $\psi_0^2 z_0^3 = 9/(8\beta)$, the denominator vanishes leading to a singularity of order $-4/3$ in addition to $-1/2$ coming from the $(1-v)^{-1/2}$ factor.

In order to proceed further and probe the behavior near the transition points, we approximate a factor which is well behaved at the singular point and put $\left( 1-\beta \psi_0^2 v \right)^{4/3} \simeq \left( 1-\beta \psi_0^2 \right)^{4/3}$ since the singularity would be located at $v=1$.
Accordingly, the integral in Eq.~\eqref{deltaEquationIntegral} can be evaluated analytically, leading to
\begin{eqnarray}
	\delta \simeq -\frac{448}{405} \, z_0^4\psi_0 \left( 1-\beta \psi_0^2 \right)^{4/3}
	{}_2 F_1 \left( 1,\frac{4}{3}; \frac{3}{2}; \frac{8}{9} \, \beta \psi_0^2 z_0^3 \right) \, , \notag 
\end{eqnarray}
where ${}_2 F_1$ denotes the hypergeometric function~\cite{abramowitz72} which, for $x \to 1$ can conveniently be approximated as
\begin{equation}
	{}_2F_1\!\left(1,\frac{4}{3};\frac{3}{2};x\right) \sim \frac{\pi^{3/2}}{\operatorname{\Gamma}(1/6) \operatorname{\Gamma}(4/3)} \, (1-x)^{-5/6} \, ,
\end{equation}
where $\operatorname{\Gamma}$ denotes the Gamma function~\cite{abramowitz72}.

The evolution of the second order parameter $\delta^{-1}$ around the transition points reads
\begin{equation}
	\delta^{-1} \sim  -\frac{\Lambda}{z_0^4 \psi_0} \left( 1-\beta \psi_0^2 \right)^{-4/3}
	\left(1- \frac{8}{9} \, \beta \psi_0^2 z_0^3\right)^{5/6} \, , 
\end{equation}
with the prefactor 
\begin{equation}
	\Lambda := \frac{405}{448} \frac{\operatorname{\Gamma}(1/6) \operatorname{\Gamma}(4/3)}{\pi^{3/2}} \, . 
\end{equation}

For a given initial distance from the wall, the transition angle is estimated as $\psiT = -3/\left(8\beta z_0^3\right)^{1/2}$ and thus
\begin{equation}
	\delta^{-1} \sim \left(\psi_0 - \psiT\right)^{5/6} \, , \label{qPrime-scaling-Theta}
\end{equation}
around the transition point, bearing in mind that $\psi_0$ and $\psiT$ are both negative quantities. 
Similar, by considering a given initial inclination $\psi_0$, the transition is expected {to occur} at a height $\zT = \frac{1}{2} \left(9/(\beta\psi_0^2)\right)^{1/3}$ and thus 
\begin{equation}
	\delta^{-1} \sim \left( \zT-z_0 \right)^{5/6} \, , \label{qPrime-scaling-Z}
\end{equation}
around the transition point.
Indeed, these scaling behaviors of the order parameters as derived from the far-field model are in a good agreement with the numerical results presented in Fig.~\ref{Fig-Esc_Both_Shifted}.

{Even though the far-field model is found to be able to capture the scaling behavior around the transition point between the escaping and trapping states, it is worth mentioning that this model nonetheless is not viable for predicting the swimming trajectories accurately.
As the swimmer gets to a finite distance close to the wall, the far-field approximation is not strictly valid.
An accurate analytical prediction of the swimming trajectories would thus require to account for the general $z$-dependence of the averaged swimming velocities and inclination.}

\section{Effect of rotation}\label{section-effect-rotation}

\subsection{State diagram}

\begin{figure}
\begin{center}
\includegraphics[scale=1]{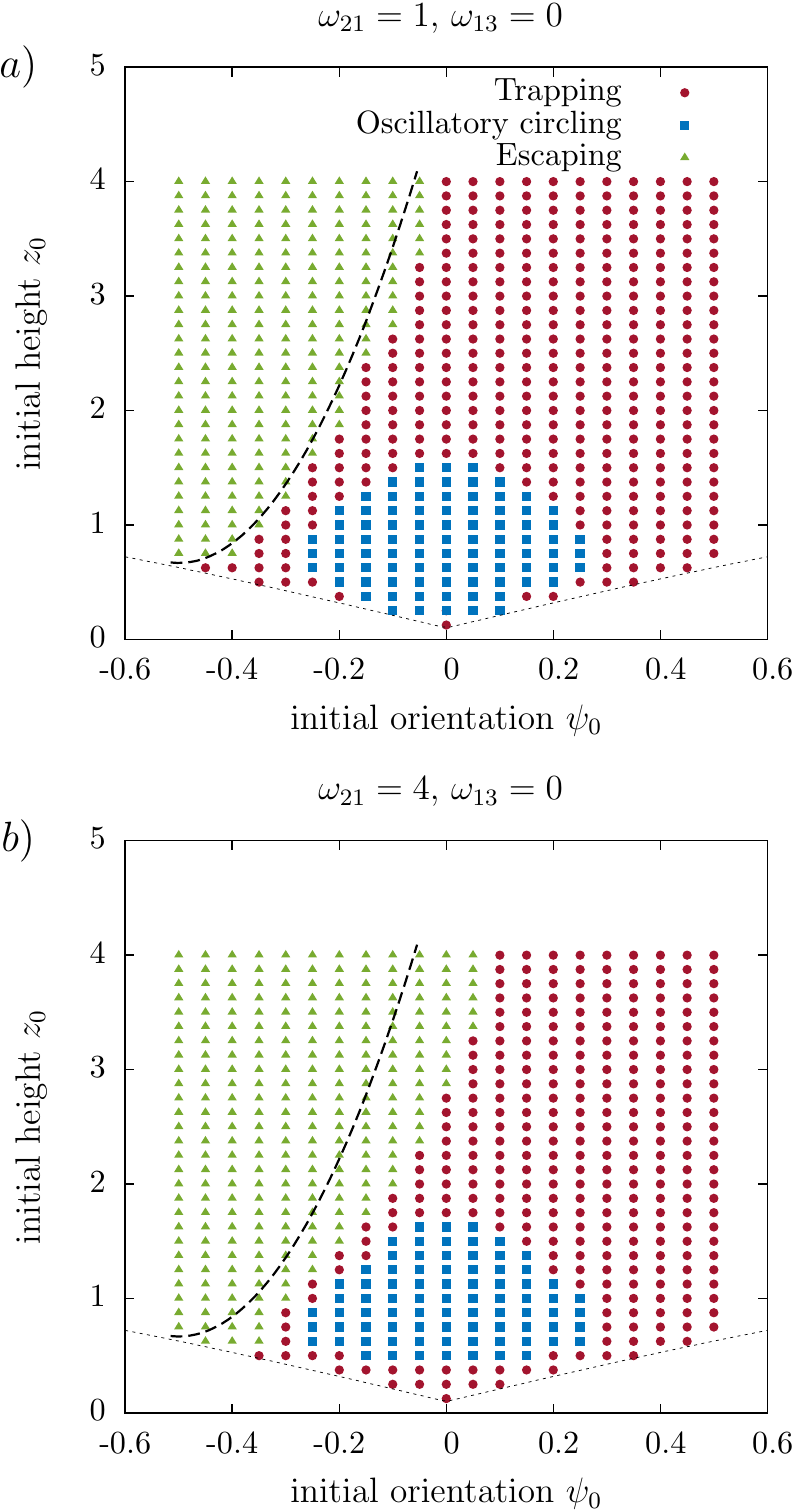}
\end{center}
\caption{(Color online) State diagram of swimming near a hard wall for a non-vanishing angular velocity along the swimming axis where $a)$ $\omega_{21}=1$ and $b)$ $\omega_{21}=4$.
Here $\omega_{13}=0$.
{The dashed line displays the boundary at the transition between the trapping and escaping states for the non-rotating system $(\omega_{21}=\omega_{13}=0).$}
The other parameters are the same as in Fig.~\ref{Phase-Diagram-Hard-Wall}.
}
\label{Phase-Diagram-Hard-Wall-3D}
\end{figure}

\begin{figure*}
\begin{center}
\includegraphics[scale=1]{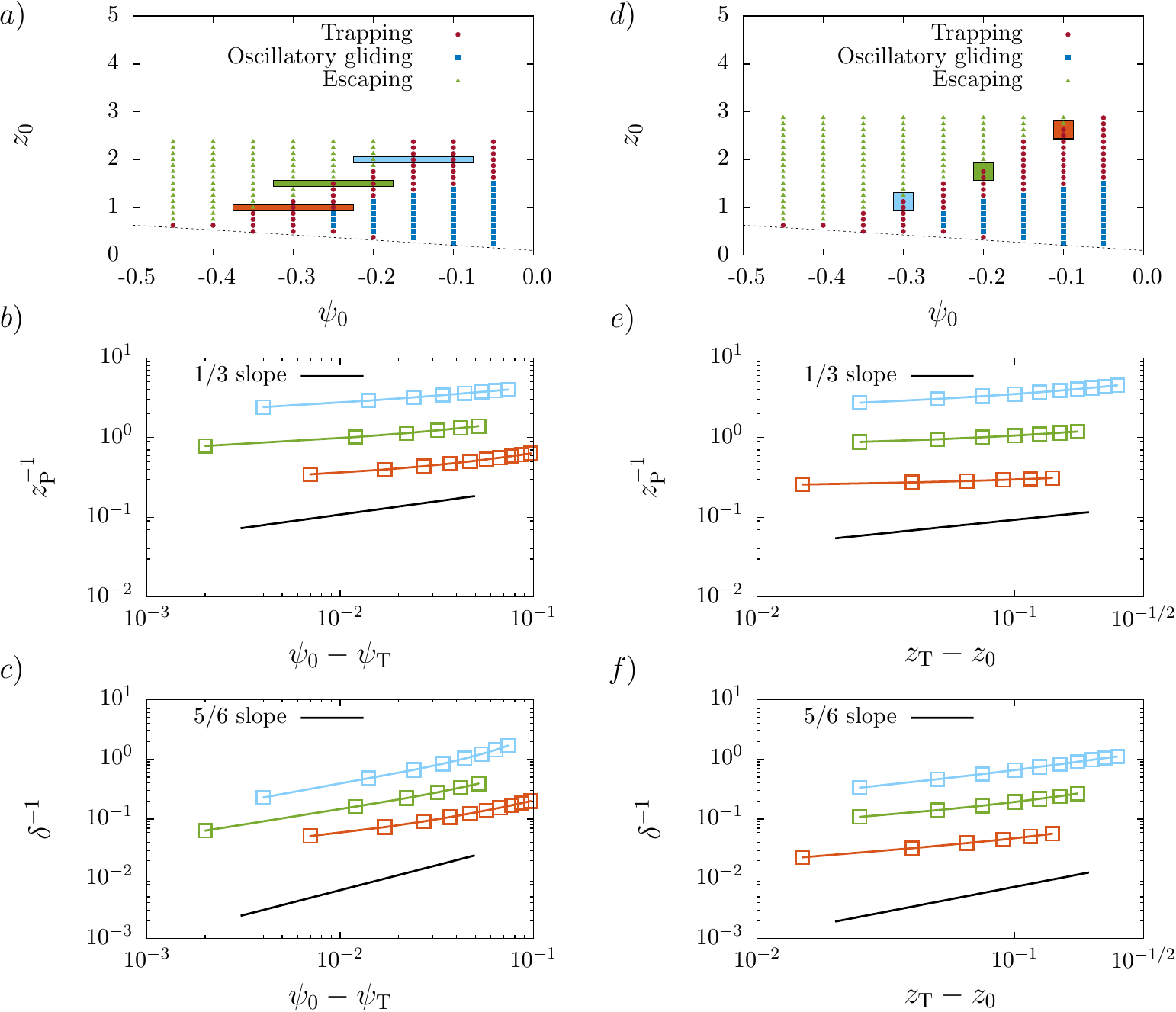}
\end{center}
\caption{(Color online) Log-log plots of the first and second order parameters at the transition point between the trapping and escaping states in the 3D case for $\omega_{21} = 1$ and $\omega_{13}=0$, as obtained from the numerical simulations. 
{The curves associated with the green and blue paths are respectively shifted for the sake of readability on the vertical scale by factors of 3 and 9.}
The solid lines are a guide for the eye.
}
\label{Fig-Esc_Both_Shifted-3D}
\end{figure*}

Having investigated the state diagram of swimming near a wall in the absence of rotation, and provided an analytical theory rationalizing our findings on the basis of a far-field model, we next consider the situation where the spheres are allowed to rotate around the swimming axis.
For flagellated bacteria, e.g., \textit{E. coli}, which swim by the action of molecular rotary motors, the flagellum undergoes counterclockwise rotation (when viewed from behind the swimmer) at speeds of $\sim 100$~Hz~\cite{lowe87, magariyama01}, whereas the cell body rotates in the clockwise direction for the bacterium to remain torque-free, at speeds of $\sim 10$~Hz~\cite{macnab77, lauga06}.
Based on these observations, we assume that the spheres 1 and 3 rotate at the same rotation rate to mimic the rotating flagellum such that  $\omega_{13}=0$, whereas the sphere 2 represents the cell body which rotates in the opposite direction.
Accordingly, $\omega_1=\omega_3 <0$ and $\omega_2>0$, and thus the relative rotation rate $\omega_{21} \equiv \omega_2-\omega_1$ has to be positive.

In Fig.~\ref{Phase-Diagram-Hard-Wall-3D}, we present the state diagram of the swimming behavior near a wall for two different values of the relative rotation rate $\omega_{21}$.
We observe that the state diagram is qualitatively similar to that obtained in the 2D case, shown in Fig.~\ref{Phase-Diagram-Hard-Wall}, where three distinct states of motion occur depending on the initial orientation and distance from the wall.
The main difference is that the oscillatory-gliding state found earlier is substituted by an oscillatory circling in the clockwise direction, at a constant mean height above the wall.
Indeed, the clockwise motion in circles has been observed experimentally for swimming \textit{E. coli} bacteria near surfaces~\cite{diluzio05} and is a natural consequence of the fluid-mediated hydrodynamics interactions with the neighboring interface and the force- and torque-free constraints imposed on the swimmer~\cite{lauga06}.

Upon increasing the rotation rate, we observe that the escaping state is enhanced to the detriment of the trapping state.
For instance, for $\omega_{13}=4$ (Fig.~\ref{Phase-Diagram-Hard-Wall-3D}~$b)$), even though the swimmer is initially pointing toward the wall an angle $\psi_0=0.05$, it can surprisingly escape the wall trapping if $z_0 \ge 3.5$.
This behavior is most probably attributed to the wall-induced hydrodynamic coupling between the translational and rotational motions, which tends to align the swimmer away from the wall.
We further observe that increasing the rotation rate favors the trapping of the swimmer if it is initially released from distance close to the wall, for $z_0<0.5$.

\subsection{Transition between states}

\subsubsection{Transition between the trapping and escaping states}

As in the 2D case, we define two relevant order parameters $\zP^{-1}$ and $\delta^{-1}$ quantifying the state transition between the trapping and escaping states.
We keep the definition of the first order parameter~$\zP^{-1}$ as the inverse of the peak hight.
By considering the 2D projection of the trajectory on the $(xy)$ plane, we define the second order parameter~$\delta^{-1}$ for the 3D motion as the inverse of the curvilinear distance along the projected path, corresponding to the occurrence of the peak.

In Fig.~\ref{Fig-Esc_Both_Shifted-3D}, we present a log-log plot of the order parameters $\zP^{-1}$ and $\delta^{-1}$ versus $\psi_0-\psiT$ (subfigures $a), b),$ and $c)$), and versus $\zT-z_0$ (subfigures $d), e),$ and $f)$) along example paths on the state diagram shown in Fig.~\ref{Phase-Diagram-Hard-Wall-3D}~$a)$, for $\omega_{21}=1$.
We observe that both order parameters exhibit analogous scaling behavior around the transition point as in the 2D case.
We will show that the general 3D case can approximatively be mapped into a 2D representational model by considering the local reference frame along the curvilinear coordinate line.
{Nevertheless, the power laws predicted analytically may not be strictly obeyed as the scaling exponents 1/3 and 5/6 derived above may not be displayed properly, notably along the vertical paths in the state diagram (Fig.~\ref{Fig-Esc_Both_Shifted-3D}~$e)$ and $f)$).
This mismatch is most probably a drawback of the simplistic approximations involved in the analytical theory proposed here for the rotating system whose derivation is outlined
in Sec.~\ref{kombinierteBewegung} below.
}

\subsubsection{Transition between the trapping and oscillatory-circling states}

We next consider the transition between the trapping and oscillatory-gliding states and define in a similar way, as in the 2D case, two relevant order parameters controlling the state transition.
{As before, we define the first order parameter as the magnitude of the scaled swimming velocity parallel to the wall averaged over one oscillation period, $\mu:= \overline{V_\parallel}/V_0$.
The second order parameter~$A$ is defined in an analogous way as the amplitude of the oscillations.  
The evolution of the order parameters have basically a similar behavior to that shown in Fig.~\ref{Fig-Osc_Mean} where the transition between the oscillatory-circling and trapping states is found also to be first order discontinuous
(see Fig.~1 in the Supporting Information for further details.)
}

In the following, we present an extension of the far-field model presented in Sec.~\ref{section-far-field-model} in order to assess the effect of the rotational motion of the spheres on the swimmer dynamics.

\begin{figure}
\begin{center}
\includegraphics[scale=1.1]{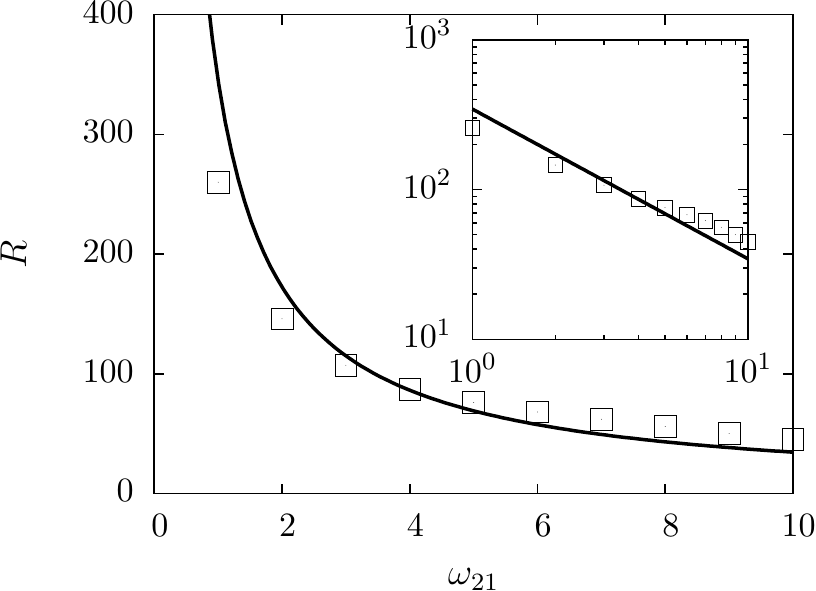}
\end{center}
\caption{Radius of curvature versus the scaled relative rotation rate $\omega_{21}$.
Solid line is the analytical prediction stated by Eq.~\eqref{radiusOfCurvatureEq} and symbols are the numerical simulations. The inset shows the same plot in a log-log scale.}
\label{Curvature}
\end{figure}

\subsection{Far-field model}

\subsubsection{Pure rotational motion}

We first consider the situation where $K=0$ and confine ourselves for simplicity to the case where the swimmer is aligned parallel to the wall for which $\psi = 0$.
The system of equations governing the swimmer dynamics at leading order in $a$ read
\begin{subequations}
	\begin{align}
		\frac{\Intd x}{\Intd t} &= -a^5 M(z) \sin\phi \, , \label{diffX_PureRot} \\
		\frac{\Intd y}{\Intd t} &= a^5 M(z) \cos\phi \, , \label{diffY_PureRot} \\
		\frac{\Intd \phi}{\Intd t} &= -a^5 Q(z) \, , \label{diffPhi_PureRot}
\\		\frac{\Intd \theta}{\Intd t}  &=0 \, ,  
	\end{align}\label{pure-rotational-motion} 
\end{subequations}
 where we have defined 
\begin{equation}
	Q(z) := \frac{\omega_{13}+2\omega_{21}}{24} 
	     	\left( \frac{1}{z^4}-\frac{z}{\xi^{5}} \right) + 2M(z)  \, , 
\end{equation}
and
\begin{equation}
	M(z) := \left( \frac{1}{24z^4} - \frac{4z}{3\zeta^5} \right) (\omega_{13} - \omega_{21})  \, , 
\end{equation}
wherein $\zeta := \left(1+4z^2\right)^{1/2}$ and $\xi := \left(1+z^2\right)^{1/2}$.
It can be seen that if $\omega_{13} = \omega_{21}$, for which the rotation rate of the central sphere is the average of the rotation rates of the spheres~2 and~3, the translational velocity vanishes and thus the swimmer undergoes a pure rotational motion around the central sphere. 
For $\omega_{13}=0$, the rotation rate $\dot{\phi}$ has a maximum values for $z \approx 0.2448$ and exhibits a decays as $z^{-6}$ in the far-field limit.

\subsubsection{Combined translation and rotation}\label{kombinierteBewegung}

We next combine the translational and rotational motions and write approximate equations governing the dynamics of the swimmer.
As can be inferred from Eqs.~\eqref{pure-rotational-motion}, the leading-order terms in the swimming velocities for a pure rotational motion scale as $a^5$.
For the translational motion $(K\ne 0)$, we have shown that at leading order, these velocities scale linearly with $a$ (c.f.~Eqs.~\eqref{approximate-equations-2D}).
Therefore, the approximated governing equations about $\psi=0$ for the combined translational and rotational motions are given by
\begin{subequations}
	\begin{align}
		\frac{\Intd x}{\Intd t} &= -  \frac{7}{24} \, aK \cos\phi \, , \label{eqDiffX1Rot} \\
		\frac{\Intd y}{\Intd t} &= -  \frac{7}{24} \, aK \sin\phi \, , \label{eqDiffY1Rot} \\
		\frac{\Intd z}{\Intd t} &= aK \left( \frac{7}{24}-\frac{21}{64} \frac{1}{z^3} \right) \psi \, , \label{eqDiffZ1Rot} \\
		\frac{\Intd \psi}{\Intd t} &= -\frac{135 }{512} \frac{aK}{z^4}   \, , \label{eqDiffTheta1Rot} \\
		\frac{\Intd \phi}{\Intd t} &=  -a^5 Q(z) \, . \label{eqDiffPhi1Rot}
	\end{align}
\end{subequations}

Defining the curvilinear coordinate $s$ along the projection of the particle trajectory on the $(xy)$ plane such that $\Intd s^2 = \Intd x^2 + \Intd y^2$, 
Eqs.~\eqref{eqDiffX1Rot} and \eqref{eqDiffY1Rot} yield
\begin{equation}
	\frac{\Intd s}{\Intd t} = -  \frac{7}{24} \, aK  \, . \label{eqDiffS1Rot} 
\end{equation}

The system of equations composed of \eqref{eqDiffZ1Rot}, \eqref{eqDiffTheta1Rot} and \eqref{eqDiffS1Rot} is mathematically equivalent to that earlier derived in the 2D case and stated by Eqs.~\eqref{approximate-equations-2D}.
In the far-field limit, the effect of the rotation of the spheres along the swimmer axis intervenes only through Eq.~\eqref{eqDiffPhi1Rot} describing the temporal change of the azimuthal angle~$\phi$.
Therefore, by appropriately redefining the second order parameter $\delta^{-1}$ as the curvilinear coordinate corresponding to the peak height, the order parameters  $\zP^{-1}$ and $\delta^{-1}$ are expected to exhibit the same scaling behavior as in the 2D case.

Finally, we calculate the radius of curvature of the swimming trajectory in the special case when $\psi_0=0$ and $z_0= 0.75$ for which the swimmer remains typically at a constant height above the wall.
According to Eq.~\eqref{eqDiffPhi1Rot}, the azimuthal angle changes linearly with time, and thus the swimmer perform a circular trajectory of radius
\begin{eqnarray}
	R = \frac{7}{24} \frac{|K|}{a^4 Q(z_0)} \sim \omega_{21}^{-1} \, , \label{radiusOfCurvatureEq}
\end{eqnarray}
for $\omega_{13}=0$.
Interestingly, the radius of curvature decays as a fourth power with $a$, while it decreases linearly with the relative angular velocity $\omega_{21}$.
Fig.~\ref{Curvature} show a quantitative comparison between analytical predictions and numerical simulations over a wide range of relative rotation rates.
While the numerical results show a slightly slower decay with $\omega_{21}$, the agreement is reasonable considering
the approximations involved in the analytical theory.

\section{Conclusions}\label{section-conclusions}

Inspired by the role of near-wall hydrodynamic interactions on the dynamics of living systems, particularly swimming bacteria\cite{lauga05} and the formation of biofilms\cite{loosdrecht1990}, we have explored the behavior of a simple model three-sphere swimmer proposed by Najafi and Golestanian \cite{najafi04} in the presence of a wall. Modeling the swimmer by three aligned spherical beads with periodically time-varying mutual distances, we have analyzed the long-time asymptotic behavior of the swimmer depending on its initial distance and orientation with respect to the wall. We have found that there are three regimes of motion, leading to either trapping of the swimmer at the wall, escape from the wall, or a non-trivial oscillatory gliding motion at a finite distance above the wall. We have found that these three states persist also when we allow the beads to rotate. The rotational motion of the beads, introduced to mimic to the rotation of a cell flagellum and a counter-rotation of its body, renders the near-wall motion of the swimmer fully three-dimensional, as opposed to the quasi-two-dimensional motion in the classic Najafi and Golestanian design. 

Having classified the swimming behavior, we have quantified the transition between different states by introducing the appropriate order parameters and measuring their scaling with the initial height and orientation. Using the far-field analytical calculations, we have shown that the scaling exponents obtained from numerical solutions of the equations of motion of the swimmer can be found exactly from the dominant asymptotic behavior of the flow field. Moreover, we have demonstrated that in the presence of internal rotation, the three-dimensional dynamics in the far-field approach can be mapped onto a quasi-two-dimensional model and thus the scalings found in both cases remain the same. We have verified the analytical predictions with numerical solutions, finding a very good agreement. This suggests that in order to grasp the general complex dynamics of the swimmer near an interface, it is sufficient to include the dominant flow field. 

In view of recent experimental realizations of the three-sphere swimmer using optical tweezers\cite{Leoni2009, grosjean16}, we hope that the findings of this paper may be verified experimentally. On one hand, it would be interesting to see the purely translational case, varying only the distances between spheres. It might prove more challenging to construct a swimmer that would actually be capable of performing an internal rotation, yet it is an exciting perspective due to the relevance of this simple model to the widely used singularity representations for swimming microorganisms near interfaces\cite{lopez14}.

\section*{Description of the supplemental information}

{
In the Supporting Information (available at [URL will be inserted by the publisher]), we provide the elements of the matrix resulting from the linear system of equations governing the generalized motion of a three-sphere swimmer near a wall given by \eqref{force-free-torque-free}, \eqref{linearEvolution_123}, \eqref{projOmegaThetaPhi} and \eqref{differenceAngVel_123}.
In addition, we provide the far-field expressions of the mobility functions used in the analytical model.
Finally, we present the evolution of the order parameters $A$ and $\mu$ in the oscillatory circling state associated with the 3D system.}

{The movies 1 and 2 illustrate a swimmer initially released from  $z_0=1$ at $\psi_0 = -0.38$ (trapping) and $\psi_0=-0.39$ (escaping). 
The movies 3 and 4 illustrate the oscillatory-gliding state for $\psi_0=0$, for a swimmer initially released from $z_0=0.75$ and $z_0=1$.
The movie 5 shows the oscillatory circling state of a swimmer initially located at $z_0=1$ above the wall, released at an angle $\psi_0=0$ for $\omega_{21}=2$ and $\omega_{13}=0$.
}


\begin{acknowledgments}
The authors gratefully acknowledge support from the DFG (Deutsche Forschungsgemeinschaft) -- SPP 1726.
This work has been supported by the Ministry of Science and Higher Education of Poland via the Mobility Plus Fellowship awarded to ML. ML acknowledges funding from the Foundation for Polish Science within the START programme. 
This article is based upon work from COST Action MP1305, supported by COST (European Cooperation in Science and Technology).
\end{acknowledgments}




%

\end{document}